\documentclass[runningheads]{llncs}
\usepackage[T1]{fontenc}

\usepackage{mathpartir}
\usepackage{eurosym}
\usepackage{amsmath}
\usepackage{amssymb,yhmath}
\usepackage{xcolor}
\usepackage{stmaryrd,textcomp}
\usepackage{mathbbol}
\usepackage{mathtools}
\usepackage{graphicx}
\usepackage{upgreek}
\usepackage{tikz}
\usepackage{tikzscale}
\usepackage{url}
\usepackage{stmaryrd}
\usepackage{fancyvrb}
\usepackage{hyperref}
\usepackage{arydshln}
\usepackage{float}
\usepackage{listings}
\usepackage{todonotes}
\usepackage{colortbl}
\usepackage{hhline}
\usepackage{multirow}
\usepackage{adjustbox}
\usepackage{array}
\usepackage{cleveref}
\usepackage{xspace}
\usepackage{tcolorbox}

\newcolumntype{R}[2]{%
    >{\adjustbox{angle=#1,lap=\width-(#2)}\bgroup}%
    l%
    <{\egroup}%
}

\newcommand{\Python}{{\sf Python}}

\renewcommand{\P}{{\tt P}}
\renewcommand{\S}{\texttt{S}}
\newcommand{\Pset}{\mathbb{P}}
\newcommand{\Cset}{\mathbb{C}}
\newcommand{\jexp}[3]{#1 \vdash #2 \blacktriangleright #3}
\newcommand{\ifte}[3]{{\tt if}\,(#1):\; #2\; {\tt else}:\; #3\;}
\newcommand{\while}[2]{{\tt while}\,(#1):\; #2\;}
\newcommand{\for}[3]{{\tt for}\; #1 \; {\tt in} \; {\tt range}(#2):\; #3\;}
\newcommand{\Var}{\mathit{Var}}

\newcommand{\deffun}[3]{{\tt def} \; #1(#2): \; #3}

\newcommand{\eqdef}{\stackrel{\textsf{\tiny def}}{=}}

\newcommand{\subst}[2]{\{^{#1}/_{#2}\}}

\newcommand{\bigfract}[2]{\frac{^{\textstyle #1}}{_{\textstyle #2}}}
\newcommand{\rulename}[1]{{\small {\sc[#1]}}}

\newcommand{\rulenamex}[1]{\mbox{\tiny [{\sc #1}]}}

\newcommand{\sem}[1]{\llbracket#1\rrbracket}

\def \mathrule #1#2#3{	\begin{array}{l} 
                       	\rulenamex{#1}
                       	\\ 
                      	 \bigfract{#2}{#3}	
                       	\end{array}
					 }

\def \mathax #1#2{\begin{array}{l} 
                  \rulenamex{#1} 
                  \\ 
                  #2
                  \end{array}
                  }

\newcommand{\newsim}{\mbox{\footnotesize $\sim$}}
\newcommand{\LRed}[1]{\mathrel{\stackrel{#1}{\Longrightarrow}}}

\newcommand{\ie}{i.e.\@\xspace}

\definecolor{prompt1}{rgb}{0.8,0.8,0.99}
\definecolor{prompt2}{rgb}{0.8,0.97,0.8}
\definecolor{prompt4}{rgb}{0.9,0.8,0.98}
\definecolor{prompt3}{RGB}{250, 205, 90}

\definecolor{classic}{rgb}{0.8,0.97,0.8}
\definecolor{notorious}{rgb}{0.97,0.97,0.8} 
\definecolor{uncommon}{rgb}{0.97,0.8,0.8} 

\definecolor{lg}{gray}{0.97}  
\definecolor{mg}{gray}{0.90}  
\definecolor{dg}{gray}{0.83}  
\definecolor{white}{gray}{1}  

\begin{document}
\title{Assessing Code Understanding in LLMs}
%
%
%

\author{
C. Laneve\inst{2}
\and
A. Spanò\inst{1}
\and
D. Ressi\inst{1}
\and  
S. Rossi\inst{1}
\and
M. Bugliesi\inst{1}
}
\authorrunning{Laneve, Spanò, Ressi et al.}
%
\institute{
DAIS, Ca' Foscari University of Venice, Italy\\
\email{\{alvise.spano,dalila.ressi,sabina.rossi,michele.bugliesi\}@unive.it}\\
\and
DISI, University of Bologna, Italy \\
\email{cosimo.laneve@unibo.it}}
\maketitle              
\begin{abstract}
We present an empirical evaluation of Large Language Models in code understanding associated with non-trivial, semantic-preserving program transformations such as copy propagation or constant folding. 
Our findings show that LLMs fail to judge semantic equivalence in approximately 41\% of cases when no context is provided and in 29\% when given a simple generic context.
To improve accuracy, we advocate integrating LLMs with code-optimization tools to enhance training and facilitate more robust program understanding.

\end{abstract}

\section{Introduction}
\label{sec:intro}

Modern Large Language Models (LLMs) exhibit remarkable capabilities in tasks related to programming, including code generation, comprehension, processing, and analysis \cite{cao2023comprehensive}.  Tools like ChatGPT \cite{chatgpturl}
, GitHub Copilot \cite{githuburl}
, and Amazon Q Developer \cite{amazonqurl} (previously known as CodeWhisperer) 
are widespread both in academy and industry to generate, optimize, and fix programs, as well as for tasks such as 
vulnerability detection and malware analysis. However, no thorough assessment appears to have been conducted on their robustness in scenarios requiring a nuanced semantics understanding of the code. In fact, while syntactic correctness and source-level transformations such as variable renaming or introduction of no-op instructions have been explored to evaluate the robustness of LLMs~\cite{pour2021search,henkel2022semantic,rabin2020evaluation}, 
their ability to grasp program equivalence remains underexamined when it comes to non-trivial semantic-preserving transformations and more general semantic equivalence questions. 

As the theoretical space of semantically equivalent programs is infinite - and determining equivalence is undecidable in the general case - we narrow our focus to a specific case of transformations that guarantee identical outputs for identical inputs. In particular, we examine source-to-source transformations inspired by standard compiler-level optimizations that enhance code performance: \emph{copy propagation} and \emph{constant folding} of complex expressions~\cite{AhoUllman}. 

Our assessment follows the methodology outlined below.
\begin{enumerate}
\item
First, we formalize copy propagation and constant folding as source-level transformations for a subset of the Python language. We target Python as widely adopted for model training and testing, and restrict to a subset to ease the formalization of the transformations. 
\item
We collect a benchmark of non-trivial use cases, and code them in the chosen Python subset: the use cases include popular algorithms of which the LLMs are likely aware, such as Fibonacci or Eratosthenes' Sieve, less popular operations on arrays and lists, and other rather uncommon operations such as anti-aliasing and type unification.
\item
We apply systematic perturbations to the source samples in the benchmark so that the original programs can be recovered by applying the source-level transformations of interest. By analyzing whether the LLMs can correctly judge the equivalence between the perturbed and original versions, we gain insight into their ability to reason about code and detect deeper relationships than superficial syntactic reorderings.
\item
We conduct our assessment on seven mainstream LLMs 
evaluating their performance based on two classes of prompts. A first, naive prompt simply asks to test the equivalence between the reference samples and their perturbed versions; a second, slightly refined prompt informs the LLMs that the perturbations may be derived by either copy propagation, constant folding, or combinations thereof, without further specifying the specific implementation of the transformations. 
\end{enumerate}

\noindent
Our findings reveal that LLMs are inaccurate in recognizing the perturbed, yet semantically equivalent, variants of programs in 41\% of the cases when no context is provided and in about 29\% of the cases when given a simple additional context. Among the different models, Anthropic Claude proves to be the best performer, with average failure rates of 19\%, while Gemini is the worst, failing in 55\% of the cases.
As a further experiment, we executed the same tests by adding to the prompt semantic-breaking perturbations. Interestingly, all LLMs performed better, somehow reinforcing the common insight that detecting a mistake is easier than proving correctness. In this case, the overall performance improves, yielding a 24\% error rate with drops to only 18\% for the contextual prompt. 

\medskip
Collectively, the reported results provide clear evidence that LLMs are still far from robust for evaluating non-trivial transformations, even when confined to specific, well-defined cases. The poor performance observed is likely due to insufficient training for the task, and additional training focused on semantically preserving code transformations is therefore necessary for improvement.
However, traditional training methods would be slow and costly, as they would require expert supervision to assess semantic equivalence across a vast code base, even for specific transformations like constant folding. To lower both the learning curve and associated costs, we advocate for integrating LLMs with automatic tools that implement code transformations, such as those proposed in this paper. Such integration would serve two complementary purposes. First, it would make the tools available for self-supervised training starting from the existing code base (simply, the tools would be employed to produce semantic equivalent transformations of the existing samples in the code base). Secondly, and more interestingly, they could be employed as general-purpose code pre-processors to eliminate ``noisy'' bits, thus enhancing LLMs in their code understanding task. 

\paragraph{Paper Structure}
The rest of the paper is structured as follows. Section~\ref{sec:related} reports
and discusses related works. Section~\ref{sec:methodology} presents in detail our methodology, namely the two code transformations (copy propagation and constant folding), 
the definition of the dataset and the prompts. Section~\ref{sec:results} reports our results and Section~\ref{sec:discussion} discusses them, analyzing possible solutions to improve reliability.
We conclude in Section~\ref{sec:conclusions} where we also present future work. The Appendix contains technical material that will be removed from the conference paper.

\section{Related Work}
\label{sec:related}
The integration of formal methods tools with AI techniques and ML models is an emerging practice in various domains \cite{urban2021review}, including data augmentation \cite{pellicer2023data}, dataset labeling \cite{ressi2024vulnerability}, network compression \cite{ressi2022neural,ressi2024compressing}, and security \cite{ressi2024ai,ressi2024vulnerability}: formal methods provide a structured approach to defining and validating semantic-preserving transformations, while AI techniques help enhance the scalability of the verification processes. 

LLM models are increasingly being evaluated in their performance with code generation tasks. In particular, 
the performance of Github Copilot, one of the most exploited applications for code generation and manipulation, has been assessed against selected fundamental algorithms such as sorting and data structure implementation~\cite{wermelinger2023using,dakhel2023github}. 

Our present focus is more directly related to the recent stream of research work that investigates the performance of LLMS in code analysis and bug detection. In particular, the results by \cite{fang2024large,dolcetti2024helping} align with our own findings, confirming that even trivial modifications, such as variable renaming, can lead to performance degradation.  On a different, but still related account, \cite{aydin2024leaking} analyzes ML-based attacks on fully homomorphic encryption via side channels, using code transformation techniques to show that unoptimized code hinders such attacks. 

However, to our knowledge, no prior work has systematically analyzed the impact of fine-grained optimizations such as the ones of our present interest. We should also mention that the analysis itself is challenging as the evaluation metrics depend on several factors, including both the varying prompting functionalities offered by the models and the often nuanced format of the responses\cite{maroengsit2019survey}.

A key distinction in chatbot evaluation is between \textit{user prompts} and \textit{system prompts}. User prompts do not require the API and include basic prompting, zero or few-shot prompting, chain-of-thought prompting, and role-based or instruction-tuned prompting. System prompts, in turn, support customization of behavior, max token limits, temperature setting, and further advanced features like function calling and logit bias, and streaming responses. \cite{sahoo2024systematic}.

Our assesment is based on user prompts rather than system prompts for two main reasons. Chatbot evaluation distinguishes between user prompts and system prompts~\cite{sahoo2024systematic}. User prompts, which do not require API access, include various prompting techniques, while system prompts enable predefined behaviors and advanced settings.
We focus on user prompts for two reasons. Firstly, free-tier chatbots, like ChatGPT, are widely used and reflect real-world interactions more accurately than open-source models like LLaMA \cite{dubey2024llama}, Gemma \cite{team2024gemma}, and Mixtral \cite{jiang2024mixtral}, which are primarily used in research. Secondly, system prompts involve hidden parameters (e.g., temperature), making results less consistent.
Additionally, we employ a zero-shot approach \cite{kojima2022large}, assessing chatbots without providing prior context or examples.

\section{Methodology}
\label{sec:methodology}
LLM evaluation is performed using Python code, as most models are well-trained in analyzing and reasoning about Python due to its simplicity and widespread use among developers. In the following subsection, we outline the techniques for code perturbation, dataset construction, and the setup of our analyses.


\subsection{Semantic-preserving Code Optimizations}
\label{sec:optimizations}

As discussed in the introduction, 
we perturb Python code in a way that requires nuanced semantic understanding. To ensure the correctness of these transformations, we leverage two well-known compiler techniques  commonly used in intermediate code optimization: 
\emph{copy propagation} of variables and \emph{constant
folding} of expressions. To apply these techniques at the source code level, we


\vspace{-0.2cm}
\begin{enumerate}
\item[(1)]
use \emph{annotations} to record information about variables 
(whether they are either constant values or copies); 
\item[(2)]
bind annotations to every control point of the code (every instruction) by means of a type inference 
system and a fixpoint analysis;
\item[(3)]
simplify the code according to the information stored in the annotations.
\end{enumerate}
The most technical aspect of this procedure is point (2), which operates as follows:
(\emph{i})
we define a set of rules that modify annotations based on their corresponding instructions;
(\emph{ii}) starting with a program where each instruction has an empty annotation, we iteratively 
infer annotations; (\emph{iii})
the inference process continues until no further modifications occur (\emph{i.e.}, a fixpoint is reached);
(\emph{iv})
once the fixpoint is reached, we proceed with the transformation step that simplifies the original 
program.
For a formal definition of point (2), the reader is referred to the Appendix, where we provide typing rules for a subset of Python statements -- sufficient to cover the codes in Section~\ref{sec:results}. The correctness of the 
overall process is beyond the scope of this paper; however, it can be proven using a similar approach to that presented in ~\cite{AhoUllman} for corresponding optimization techniques applied to intermediate code.

The two transformations are defined below. We use the following notations: given a Python program {\P},
let $\Var(\P)$ be the set of variables of $\P$ ($\Var(\cdot)$ also applies to statements $S$, with the same 
meaning); if 
$S$ is either a statement or an expression, we write $S \subst{y}{x}$ the 
substitution of every occurrence of $x$ in $S$ with $y$ (we assume that $x$ and $y$ are not redefined in 
$S$ -- they are \emph{free}).

\subsubsection{Copy Propagation.}

The annotations of the copy propagation analysis are sets $\Pset$ of pairs $(x,y)$, 
with $x, y \in \Var(\P)$, that are closed by symmetry and transitivity.
We always abbreviate $\{ (x,y), (y,x)\}$ with $x \newsim y$ whose meaning is 
``\emph{$x$ and $y$ store the same value -- they are copies}''.
The inference system, that we report in the Appendix (\emph{c.f.}~Table~\ref{tab:typeinferenceCP}),
allows us to derive judgments $\jexp{\Pset}{S}{\Pset'}$ to be read as follows: 
\emph{if the statement $S$ 
is evaluated in a memory where variables in $\Pset$ are copies then its evaluation terminates in a 
memory whose variables in $\Pset'$ are copies}. As an example, the topmost code in 
Figure~\ref{fig.cpexample}
reports the annotations we derive with the system of the Appendix (instead of writing 
$\jexp{\Pset}{S}{\Pset'}$, we write $S \; \; \Pset, \Pset'$).

In the following transformations, with an abuse of notation, we use semicolons to separate an assignment 
from its continuation. Therefore, the Python code $x = E \; \; S$ becomes $x = E \; \texttt{;} \; S$.
The code transformations $\LRed{\tt CP}$ due to copy propagation analysis are the
following ones:
\begin{enumerate}
\item
if $\jexp{\Pset}{x = y}{\Pset'}$ with $x \newsim y \subseteq \Pset$ then 
\[
 x = y  \qquad \LRed{\tt CP} \qquad \varepsilon
\]
(in this case $\Pset = \Pset'$: we are erasing $x = y$);
\item
if $\jexp{\Pset}{x = y}{\Pset'}$ and $\jexp{\Pset'}{S}{\Pset''}$, such that either (\emph{i}) every annotation 
in the proof of $\jexp{\Pset'}{S}{\Pset''}$ contains $x \newsim y$ or (\emph{ii}) $x \notin \Var(S)$, then 
\[
x = y \; \texttt{;} \; S \qquad \LRed{\tt CP} \qquad  
S\subst{y}{x} \;\texttt{;}  \; x = y
\]

\item
if $\jexp{\Pset}{y = e \; \texttt{;} \; S  \; \texttt{;} \; x = y}{\Pset'}$ 
and  $x \newsim y \subseteq \Pset$ and $x \notin \Var(S)$ then
\[
y = e \; \texttt{;} \; S \; \texttt{;} \; x = y 
\qquad \LRed{\tt CP} \qquad
x = e \; \texttt{;} \; S\subst{x}{y} \; \texttt{;} \; y =x  
\]
(in this case $x \newsim y \subseteq \Pset'$)

\item
if $\jexp{\Pset}{\while{e}{\!(S\; \texttt{;} \; x = y)}}{\Pset'}$ and $(x,y) \in
\Pset$ and $x \notin \Var(\S)$ then 
\[
\while{e}{\!\!(S \; \texttt{;} \; x = y)} \qquad \LRed{\tt CP} \qquad 
(\while{e\subst{y}{x}}{\!\!S}) \; \texttt{;} \; x = y 
\]
\item
if $\jexp{\Pset}{\for{i}{e}{ (S\; \texttt{;} \; x = y)}}{\Pset' }$ and $x \newsim y
\subseteq \Pset$ and $x \notin \Var(S)\setminus i$ then 
\[
\for{i}{e}{\!\!(S \; \texttt{;} \; x = y)} 
 \quad \LRed{\tt CP} \quad
(\for{i}{e\subst{y}{x}}{\!\!S}) \;\texttt{;} \; x = y 
\]
\end{enumerate}
For example, in Figure~\ref{fig.cpexample}, we show the code transformation of a simple iterative program computing the factorial.
\begin{figure}[t]
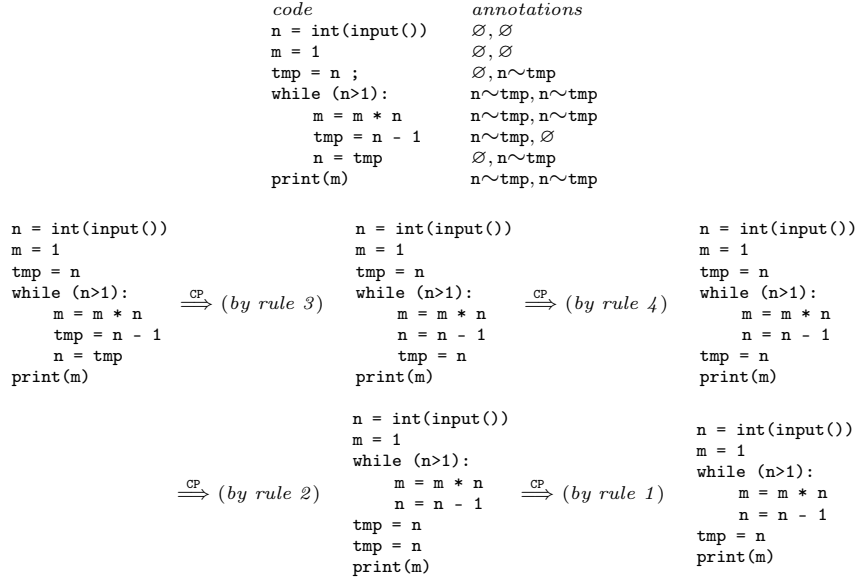

\begin{tabular}{c}
{\scriptsize
\begin{tabular}{l@{\qquad}l}
\emph{code} & \emph{annotations}
\\
{\tt n = int(input())} & $\varnothing, \varnothing$
\\
{\tt m = 1} & $\varnothing, \varnothing$
\\
{\tt tmp = n ;} & $\varnothing, {\tt n} \newsim {\tt tmp}$
\\
{\tt while (n>1):}& ${\tt n} \newsim {\tt tmp}, {\tt n} \newsim {\tt tmp}$
\\
\qquad {\tt m = m * n}& ${\tt n} \newsim {\tt tmp}, {\tt n} \newsim {\tt tmp}$
\\
\qquad {\tt tmp = n - 1}& ${\tt n} \newsim {\tt tmp}, \varnothing$
\\
\qquad {\tt n = tmp}& $\varnothing, {\tt n} \newsim {\tt tmp}$
\\
{\tt print(m)}& ${\tt n} \newsim {\tt tmp}, {\tt n} \newsim {\tt tmp}$
\end{tabular} }
\\
\\
{\scriptsize
\begin{tabular}{ll}
\begin{tabular}{l}
{\tt n = int(input())}
\\
{\tt m = 1} 
\\
{\tt tmp = n } 
\\
{\tt while (n>1):}
\\
\qquad {\tt m = m * n}
\\
\qquad {\tt tmp = n - 1}
\\
\qquad {\tt n = tmp}
\\
{\tt print(m)}
\end{tabular}
$\LRed{\tt CP}$ (\emph{by rule 3}) \quad 
\begin{tabular}{l}
{\tt n = int(input())}
\\
{\tt m = 1} 
\\
{\tt tmp = n } 
\\
{\tt while (n>1):}
\\
\qquad {\tt m = m * n}
\\
\qquad {\tt n = n - 1}
\\
\qquad {\tt tmp = n}
\\
{\tt print(m)}
\end{tabular}
$\LRed{\tt CP}$ (\emph{by rule 4}) \quad 
\begin{tabular}{l}
{\tt n = int(input())}
\\
{\tt m = 1} 
\\
{\tt tmp = n } 
\\
{\tt while (n>1):}
\\
\qquad {\tt m = m * n}
\\
\qquad {\tt n = n - 1}
\\
{\tt tmp = n}
\\
{\tt print(m)}
\end{tabular}
\\
\\
\qquad \qquad \qquad \qquad $\LRed{\tt CP}$ (\emph{by rule 2}) \quad 
\begin{tabular}{l}
{\tt n = int(input())}
\\
{\tt m = 1} 
\\
{\tt while (n>1):}
\\
\qquad {\tt m = m * n}
\\
\qquad {\tt n = n - 1}
\\
{\tt tmp = n } 
\\
{\tt tmp = n}
\\
{\tt print(m)}
\end{tabular}
$\LRed{\tt CP}$ (\emph{by rule 1}) \quad 
\begin{tabular}{l}
{\tt n = int(input())}
\\
{\tt m = 1} 
\\
{\tt while (n>1):}
\\
\qquad {\tt m = m * n}
\\
\qquad {\tt n = n - 1}
\\
{\tt tmp = n}
\\
{\tt print(m)}
\end{tabular}
\end{tabular}
}
\end{tabular}
\caption{\label{fig.cpexample} Annotations and transformations of copy propagation.} 
\end{figure}
At each step, we indicate the transformation rule that has been applied.
In particular, in the first step, we apply rule 3 when $S$ is empty. 
It is worth to notice that
since  
every transformation step modifies the program, the corresponding annotations must also be updated.
However, rerunning the inference system at every step is not necessary; instead, annotation updates can be incorporated directly within the transformation steps.  For simplicity, this aspect has been omitted in the previous discussion.
Finally, we notice that it is possible to use the standard garbage collection rule:
\begin{center}
If $\P = S \; \texttt{;} \; x = e \; \texttt{;} \; S'$  and $x \notin \Var(S')$ then remove $x = e$.
\end{center}
For example, in the last code of Figure~\ref{fig.cpexample}, we can delete the statement
{\tt tmp = n}.

\subsubsection{Constant Folding.} 

The annotations for the constant folding are \emph{abstract memories} $\Cset$, which are sets of pairs 
$(x, \mu)$ where $x \in \Var(\P)$ (for some 
Python program $\P$) and $\mu$ is either a constant value or $\top$, where $\top$ represents 
a non-constant value. The pair $(x, {\tt k})$, where {\tt k} is a constant, means that the variable $x$ 
has value {\tt k} (in the memory); $(x, \top)$ means that the variable $x$ 
is undetermined (in the memory) -- we are not able to determine its value at static time --
or \mbox{\tt ?} that represents an error.
The inference system that we report in the Appendix 
(\emph{c.f.}~Table~\ref{tab:typeinferenceCF}) allows us to derive judgments of the form $\jexp{\Cset}{S}{\Cset'}$ whose
meaning is: \emph{if $S$ is evaluated in an abstract memory where variables have the values stored in 
$\Cset$ then
its evaluation terminates with a memory whose variables have the values stored in $\Cset'$}.
The topmost code in 
Figure~\ref{fig.cfexample}
reports the annotations we derive with the system of the Appendix (as before, instead of writing 
$\jexp{\Cset}{S}{\Cset'}$, we write $S \; \; \Cset, \Cset'$).

The transformation steps for the constant folding use an auxiliary function $\sem{E}_{\Cset}$ that, given an expression $E$, evaluates $E$ in the abstract memory $\Cset$; if some variable in $E$ has 
either no value or it is \mbox{\tt ?} in $\Cset$ then the result is 
\mbox{\tt ?} (error), otherwise (some variable is  
$\top$ 
in $\Cset$ and no variable is \mbox{\tt ?}) the result is $\top$.
We use \emph{expression contexts}, written $\mathcal{E}[\;]$, which are expressions with one 
hole $[\;]$ in correspondence of a sub-expression. For example $x + [\;]$ is an expression context.
\begin{enumerate}
\item
if $S$ is either $x = \mathcal{E}[E]$ or $\ifte{\mathcal{E}[E]}{S'}{S''}$ or 
$\while{\mathcal{E}[E]}{S'}$ or $\for{i}{\mathcal{E}[E]}{S'}$ or $f(\cdots, \mathcal{E}[E], \cdots)$
and 
$\jexp{\Cset}{S}{\Cset'}$ 
with $\sem{E}_{\Cset}= {\tt k}$
and ${\tt k} \neq \top$ then $S$ may be rewritten as follows: 
\[
\begin{array}{rcl}
x = \mathcal{E}[E] & \quad \LRed{\tt CF} \quad & x = \mathcal{E}[{\tt k}]
\\
\ifte{\mathcal{E}[E]}{S'}{S''} & \quad \LRed{\tt CF} \quad & \ifte{\mathcal{E}[{\tt k}]}{S'}{S''}
\\
\while{\mathcal{E}[E]}{S'} & \quad \LRed{\tt CF} \quad & \while{\mathcal{E}[{\tt k}]}{S'}
\\
\for{i}{\mathcal{E}[E]}{S'} & \quad \LRed{\tt CF} \quad & \for{i}{\mathcal{E}[{\tt k}]}{S'}
\\
f(\cdots, \mathcal{E}[E], \cdots) & \quad \LRed{\tt CF} \quad & f(\cdots, {\tt k}, \cdots) 
\end{array}
\]
\item
if $\jexp{\Cset}{x = E}{\Cset'}$ 
and $\sem{E}_{\Cset}= {\tt k}$ (${\tt k} \neq \top$) and $(x,{\tt k}) \in \Cset$ then 
\[
x = E \qquad \LRed{\tt CF} \qquad \varepsilon
\]
(remove the assignment);
\item
if $\jexp{\Cset}{x = E \; \texttt{;} \; S}{\Cset'}$ 
and $x \notin \Var(S)$  then 
\[
x = E \; \texttt{;} \; S \qquad \LRed{\tt CF} \qquad S \;\texttt{;}\; x = E
\]
if $\jexp{\Cset}{x = E \; \texttt{;} \; x= E'}{\Cset'}$ 
and $x \notin \Var(E')$  then 
\[
x = E \;\texttt{;} \; x = E' \qquad \LRed{\tt CF} \qquad  x = E'
\]

\item
if either
$\jexp{\Cset}{\ifte{E}{S}{S'}}{\Cset'}$ 
or 
$\jexp{\Cset}{\while{E}{S}}{\Cset'}$
or
$\jexp{\Cset}{\for{i}{E}{S}}{\Cset'}$
then
\[
\begin{array}{rcll}
\ifte{E}{S}{S'} & \quad \LRed{\tt CF} & \quad S & (\text{when } \; \sem{E}_{\Cset}= {\it true})
\\
\ifte{E}{S}{S'} & \quad \LRed{\tt CF} & \quad S' & (\text{when } \; \sem{E}_{\Cset}= {\it false})
\\
\while{E}{S} & \quad \LRed{\tt CF} & \quad \varepsilon & (\text{when } \; \sem{E}_{\Cset}= {\it false})
\\
\for{i}{E}{S} & \quad \LRed{\tt CF} & \quad \varepsilon & (\text{when } \; \sem{E}_{\Cset} = {\tt k}
\; \text{ and } \;  {\tt k} \leq 0 )
\end{array}
\]
\end{enumerate}

In Figure~\ref{fig.cfexample}, we apply the constant folding transformations to a simple iterative program computing the factorial.
\begin{figure}[t]
\begin{tabular}{c}
{\scriptsize
{\tt \begin{tabular}{l@{\qquad}l}
\mbox{${\it code}$} & \mbox{${\it annotations}$}
\\
n = int(input()) & $\emptyset , \; 
\{ ({\tt n}, \top) \}$
\\
tmp = 1  & $\{ ({\tt n}, \top) \} , \; 
\{ ({\tt n}, \top), ({\tt tmp}, 1) \}$
\\
m = 2*tmp - 1 & $\{ ({\tt n}, \top), ({\tt tmp}, 1) \} , \; 
\{ ({\tt n}, \top), ({\tt m}, 1), ({\tt tmp}, 1) \}$
\\
while (n > tmp): & $\{ ({\tt n}, \top), ({\tt m}, \top), ({\tt tmp}, 1) \} , \; 
\{ ({\tt n}, \top), ({\tt m}, \top), ({\tt tmp}, 1) \}$
\\
\qquad  tmp = tmp + 1 & $\{ ({\tt n}, \top), ({\tt m}, \top), ({\tt tmp}, 1) \} , \; 
\{ ({\tt n}, \top), ({\tt m}, \top), ({\tt tmp}, 2) \}$
\\
\qquad m = m * n & $\{ ({\tt n}, \top), ({\tt m}, \top), ({\tt tmp}, 2) \} , \; 
\{ ({\tt n}, \top), ({\tt m}, \top), ({\tt tmp}, 2) \}$
\\
n = n - 1 & $\{ ({\tt n}, \top), ({\tt m}, \top), ({\tt tmp}, 2) \} , \; 
\{ ({\tt n}, \top), ({\tt m}, \top), ({\tt tmp}, 2) \}$
\\
\qquad tmp = tmp - 1 & $\{ ({\tt n}, \top), ({\tt m}, \top), ({\tt tmp}, 2) \} , \; 
\{ ({\tt n}, \top), ({\tt m}, \top), ({\tt tmp}, 1) \}$
\\
print(m) & $\{ ({\tt n}, \top), ({\tt m}, \top), ({\tt tmp}, 1) \} , \; 
\{ ({\tt n}, \top), ({\tt m}, \top), ({\tt tmp}, 1) \}$
\end{tabular} }}
\\
\\
{\scriptsize
\begin{tabular}{ll}
{\tt \begin{tabular}{l}
n = int(input())
\\
tmp = 1 
\\
m = 2*tmp - 1
\\
while (n > tmp):
\\
\qquad  tmp = tmp + 1
\\
\qquad m = m * n
\\
\qquad n = n - tmp + 1
\\
\qquad tmp = tmp - 1
\\
print(m)
\end{tabular}}
$\LRed{\tt CF}$ (\emph{by rule 1}) \; 
{\tt \begin{tabular}{l}
n = int(input())
\\
tmp = 1 
\\
m = 1
\\
while (n > 1):
\\
\qquad  tmp = 2
\\
\qquad m = m * n
\\
\qquad n = n - 1
\\
\qquad tmp = 1
\\
print(m)
\end{tabular}}
$\LRed{\tt CF}$ (\emph{by rule 3}) \; 
{\tt \begin{tabular}{l}
n = int(input())
\\
tmp = 1 
\\
m = 1
\\
while (n > 1):
\\
\qquad m = m * n
\\
\qquad n = n - 1
\\
\qquad  tmp = 2
\\
\qquad tmp = 1
\\
print(m)
\end{tabular}}
\\
\\
\qquad \qquad \qquad $\LRed{\tt CF}$ (\emph{by rule 3}) \; 
{\tt \begin{tabular}{l}
n = int(input())
\\
tmp = 1 
\\
m = 1
\\
while (n > 1):
\\
\qquad m = m * n
\\
\qquad n = n - 1
\\
\qquad tmp = 1
\\
print(m)
\end{tabular}}
$\LRed{\tt CP}$ (\emph{by rule 2}) \; 
{\tt \begin{tabular}{l}
n = int(input())
\\
tmp = 1 
\\
m = 1
\\
while (n > 1):
\\
\qquad m = m * n
\\
\qquad n = n - 1
\\
print(m)
\end{tabular}}
\end{tabular}
}
\end{tabular}
\caption{\label{fig.cfexample} Annotations and transformations of constant folding} 
\end{figure}

In particular, in the first step, we apply rule 1, thus computing all the constant subexpressions.
As for copy propagation, it is possible to collect garbage variables not used in the continuations.
Therefore, the assignment {\tt tmp = 1} in the last code may be removed.

\subsection{Dataset Construction}
\label{sec:dataset-construction}

The dataset comprises multiple semantically equivalent and non-equivalent variants of small Python programs, designed to assess the model’s ability to distinguish correct from incorrect implementations. We crafted 11 distinct programs in total, each available in 8 variants. These programs consist of standalone functions implementing various algorithms, differing in length, complexity, and notoriety. Some algorithms, such as Bubble Sort and the Sieve of Eratosthenes, are well known, while others, like type unification and 3D point rotation, are less commonly recognized. The dataset includes both algorithms that manipulate data structures, such as lists and arrays, and those performing mathematical computations, ensuring a diverse set of challenges. \Cref{tab:snippets} provides an overview of all included algorithms.  

\begin{table}[h]
    \centering
    \setlength{\tabcolsep}{6pt} 
\renewcommand{\arraystretch}{1.5} 
     \resizebox{0.7\textwidth}{!}{
    \begin{tabular}{|m{4cm}|m{4cm}|m{4cm}|}
    \hline 
    \textbf{Arithmetics}    & \textbf{List Manipulation}    & \textbf{Other}
    \\ \hline 
    \cellcolor{classic} Fibonacci               & \cellcolor{notorious} Remove Duplicates             & \cellcolor{uncommon} Anti Aliasing          
    \\ \hline 
    \cellcolor{classic} Primality Test          & \cellcolor{notorious} Find Occurrence               & \cellcolor{uncommon} Unification (MGU)         
    \\ \hline 
    \cellcolor{notorious} Rotate 3D Point      & \cellcolor{notorious} Count Occurrences             & 
    \\ \hline 
    \cellcolor{uncommon} Fast Fourier Transform & \cellcolor{classic} Sieve of Eratosthenes & 
    \\ \hline 
    \end{tabular}}
    \vspace{1em}
    \caption{Our dataset of Python programs. Colors mark the notoriety of the algorithm: green denotes 
    classic algorithms, well-known to LLMs; yellow marks somewhat notorious functions that are typically found in libraries; red refers to uncommon algorithms that belong to specific domains, such as computer graphics or compiler construction, and are less likely to prevail in training sets.}
    \label{tab:snippets}
\vspace{-.5cm}    
\end{table}
Each algorithm in the dataset follows a structured design, with implementations grouped into two classes:  

\begin{itemize}
    \item \textbf{Correct Implementations (Semantically Equivalent):}
    \begin{itemize}
        \item One standard implementation serving as a reference.
        \item Three perturbed versions that preserve the original semantics while applying simple transformations - copy propagation, constant folding, and a combination of both. These transformations are reversible using standard optimization rules (see \Cref{sec:optimizations}).
    \end{itemize}

    \item \textbf{Incorrect Implementations (Semantically Non-equivalent):}
    \begin{itemize}
        \item One version introducing a semantic-breaking bug in the reference implementation.
        \item Three additional versions derived from each of the perturbed correct implementations by introducing a bug. These errors involve small but meaningful changes, such as index modifications or incorrect assignments, making them semantically incorrect.
    \end{itemize}
\end{itemize}

To further challenge model recognition, especially for less common algorithms, we introduced a degree of obfuscation. Function names were anonymized, and variable identifiers were replaced with short, uninformative labels to prevent reliance on superficial patterns.

The dataset is publicly available on GitHub \cite{repository}.
Each source file contains one algorithm in the 8 variants, plus some extra code that puts the actual semantic equivalence to the test by performing unit tests for a reasonable range of inputs and comparing the outputs.

\subsection{Experimental Setup} 
\label{sec:experimental-setup}

Our experiments involved collecting and analyzing responses from several mainstream chatbots.
We formulated our case study as a multi-instance binary classification problem, where both the reference implementation and its perturbed variants were presented within the same prompt.
For each algorithm in the dataset, we designed four zero-shot prompts as follows:

\begin{center}
\setlength{\tabcolsep}{6pt} 
\renewcommand{\arraystretch}{1.5} 
     \resizebox{0.6\textwidth}{!}{
\begin{tabular}{|c|c|c|}
\cline{2-3}
\multicolumn{1}{c|}{}   & \textbf{Contextless Preamble}  & \textbf{Contextual Preamble} \\
\hline
\textbf{Single-Class}  & \cellcolor{prompt1} Prompt 1   & \cellcolor{prompt2} Prompt 2  \\
\hline
\textbf{Multi-Class}   & \cellcolor{prompt3} Prompt 3   & \cellcolor{prompt4} Prompt 4 \\
\hline
\end{tabular}}
\end{center}

\begin{description}
\item[{Prompt 1}](single-class, contextless): a contextless brief question, followed by the four correct versions of the algorithm (one reference and three perturbed);
\item[{Prompt 2}](single-class, contextual): a contextualized preamble, followed by the same four correct versions;
\item[{Prompt 3}](multi-class, contextless): Prompt 1 plus the four incorrect implementations, resulting in a total of eight code snippets that mix correct and incorrect versions;
\item[{Prompt 4}](multi-class, contextual): similar to Prompt 3, but including the contextual preamble from Prompt 2.
\end{description}

Prompt~1 is intended to ask: \textit{``Can the chatbot recognize correct implementations based solely on structure, without contextual guidance?''}  
It presents a brief, contextless question alongside four correct implementations—the reference and three perturbed versions—testing whether the chatbot can assess correctness purely from code patterns.  
Prompt~2 adds background information to evaluate whether context improves classification accuracy.  
Prompt~3 shifts focus to: \textit{``Can the chatbot distinguish between correct and incorrect implementations with no contextual support?''}  
Unlike Prompt~1, it includes both valid and flawed versions, requiring the model to rely solely on its understanding of correctness.  
Prompt~4 combines this challenge with additional context, assessing whether more information aids or complicates classification.

\begin{figure}[t]
\textbf{Contextless Preamble}
    \begin{tcolorbox}[colback=gray!20, colframe=gray!50, width=\textwidth, boxrule=0.5pt]
    \ttfamily 
    {\scriptsize Are the following functions semantically equivalent to the first one?}
    \end{tcolorbox}
\vspace{1em}
\textbf{Contextual Preamble}
        \begin{tcolorbox}[colback=gray!20, colframe=gray!50, width=\textwidth, boxrule=0.5pt]
    \ttfamily 
    {\scriptsize You are a chatbot for comparing the semantics of small Python programs.
    I will provide you with multiple implementations of the same Python function.
    The first function is the reference version. 
    The other functions are perturbed with copy propagation, constant folding or a combination of the two.
    Tell me whether the functions are semantically equivalent to the reference version or not.}
    \end{tcolorbox}
    \caption{Difference between the preamble of the contextless and contextual prompts.}
    \label{fig:preamble-prompt}
\end{figure}

Each response to the prompts was carefully reviewed by two of the authors and framed as a binary multi-instance classification problem. Specifically, each algorithm within a prompt was compared to the first function provided, serving as the reference, to assess semantic equivalence.  
The preamble or contextual priming at the beginning of the prompts follows two distinct formats, illustrated in Figure~\ref{fig:preamble-prompt}. Both preambles are zero-shot, as no labeled examples are provided.  
The contextual preamble combines multiple strategies:  
\begin{itemize}
    \item \textbf{Role prompting}: ``\textsf{You are a chatbot for comparing the semantics of small Python programs.}''
    \item \textbf{Instruction prompting}: ``\textsf{I will provide you with multiple implementations of the same Python function. The first function is the reference version.}''
    \item \textbf{Contextual information}: ``\textsf{The other functions are perturbed with copy propagation, constant folding, or a combination of the two.}''
\end{itemize}

Each of the four prompts was repeatedly submitted to seven chatbots over 10 rounds, with a new chat session initialized for each iteration.  
This resulted in a total of $11 \cdot 4 \cdot 7 \cdot 10 = 3080$ outputs.  
Chatbot responses tend to be highly verbose. To encourage the models' reasoning process, we did not request a simple yes/no answer, as we observed a performance drop in such cases.  
Each response was carefully analyzed, taking into account potential contradictions in the text and other logical inconsistencies.  
Rather than post-processing the answers using an additional AI, this preliminary series of experiments relied on human evaluation to mitigate the risk of misinterpreting verbose and complex outputs.

\Cref{tab:chatbots} enlists the chatbots put to the test along with the implemented model at the time.
The list does not include premium subscriptions except for Copilot, which requires one.
No API has been used: all experiments have been conducted manually through the web interface, when available, or through the free-to-use Visual Studio Code plugin when that was the only option (as in the case of Copilot and Amazon Q).
This approach is fundamental to our preliminary study, as we aim to evaluate chatbot performance under the same conditions in which most casual users interact with them.


\begin{table}[t]
    \centering
    \setlength{\tabcolsep}{3pt} 
    \renewcommand{\arraystretch}{1.1} 
    \resizebox{0.8\textwidth}{!}{
    \begin{tabular}{|m{4.2cm}|m{2.3cm}|m{4.2cm}|}
    \hline
    \textbf{Company} & \textbf{Product} & \textbf{Model} \\
    \hline
    \rowcolor{lg} Github & Copilot Pro & Claude 3.5 Sonnet \\
    Github & Copilot Pro  & GPT-4o \\
    \rowcolor{lg} Google & Gemini & Gemini 2.0 Flash \\
    OpenAI & ChatGPT & GPT-4o \\
    \rowcolor{lg} DeepSeek & DeepSeek & DeepSeek-R1 V3 \\
    Amazon Web Services & Q Developer & Various \\
    \rowcolor{lg} Anthropic & Claude & Claude 3.5 Sonnet \\
    \hline
    \end{tabular}
    }
    \vspace{1em}
    \caption{Chatbot models and versions involved in our experimentation as of January-February 2025.}
    \label{tab:chatbots}
    \vspace{-.7cm}
\end{table}

It is worth mentioning that Copilot serves as a frontend for multiple models; our tests focus on two: ChatGPT and Claude Sonnet.  
Despite using the same underlying technology as their standalone versions, results on Copilot differ, warranting their inclusion in our experiments.  
Similarly, Amazon Q reportedly employs Claude 3.5 Sonnet\footnote{As of February 2025, the Amazon Q Developer official blog states that Claude Sonnet is the primary model for code review and generation: \url{https://aws.amazon.com/it/blogs/devops/amazon-q-developer-inline-chat}}, alongside custom models selected based on the prompt type.  
However, both Amazon Q and Copilot Claude yield results distinct from their official counterparts.  

Additionally, chat-based interaction prevents users from adjusting the \textit{temperature} parameter \cite{davis2024temperature}, which controls response variability.  
Typically ranging from 0 to 1, a temperature of 0 ensures deterministic outputs, while higher values introduce randomness.  
Most chatbots default to ~0.7, adjusting dynamically, though the exact settings are not always disclosed. 
Consequently,  responses may vary across sessions, times, and even accounts.  To account for this variability, our experiments were conducted across different time frames and conditions to ensure a fair sampling of model behavior.

\section{Experimental Results}
\label{sec:results}

Our results are based on 3080 responses from the seven chatbots examined. For the single-class prompts (Prompt 1 and Prompt 2) we collected 4620 responses, while for the multi-class experiments (Prompt 3 and Prompt 4) we extracted 10780 evaluations, for a total of 15400 fine-grained \emph{yes/no} answers. The results are aggregated in \Cref{tab:accuracy-correct-only,tab:accuracy-prompt3_4}.

\Cref{tab:accuracy-correct-only} shows the accuracy of the class of semantically equivalent programs across all prompts for each chatbot. The rightmost column shows the average accuracy for each prompt across all chatbots. The bottom row represents the average for each chatbot across all prompts. Prompts 3 and 4 report the accuracy of the class of semantically equivalent programs, despite including both correct and incorrect code. At first glance, Claude Sonnet and DeepSeek emerge as the top performers, with Claude holding a slight edge on average across all prompts (79.09\% vs. 76.88\%). Conversely, Gemini exhibits the lowest scores in most compartments (43.72\% overall), followed somewhat unexpectedly by Copilot Claude (60.19\%). Interestingly, despite both purportedly using Claude Sonnet, the model behind Copilot appears to behave differently from Anthropic’s implementation. The rest of the models fall in the middle range, with ChatGPT surprisingly not exceeding the average. It is clear that contextual prompts, i.e., Prompts 2 and 4, enhance the accuracy of all chatbots, with the notable exception of Gemini, whose performance unexpectedly declines. The most significant improvement is observed in Copilot ChatGPT and Amazon Q, which appear to benefit greatly from contextual information. Conversely, top-performing models such as DeepSeek and Claude Sonnet show only a minor accuracy gain.

\newcolumntype{C}[1]{>{\centering\arraybackslash}p{#1}}
\begin{table}[t]
    \centering
    \setlength{\tabcolsep}{2.5pt} 
    \renewcommand{\arraystretch}{1.8} 
\resizebox{\textwidth}{!}{
        \begin{tabular}{|p{4.5cm}|>{\raggedleft\arraybackslash}p{1.8cm}|>{\raggedleft\arraybackslash}p{1.8cm}|>{\raggedleft\arraybackslash}p{1.8cm}|>{\raggedleft\arraybackslash}p{1.8cm}|>{\raggedleft\arraybackslash}p{1.8cm}|>{\raggedleft\arraybackslash}p{1.8cm}|>{\raggedleft\arraybackslash}p{1.8cm}|>
    {\raggedleft\arraybackslash}p{1.8cm}|}
    \cline{2-4}

    \multicolumn{1}{c|}{} & \multicolumn{3}{c|}{\textbf{VSCode Plugin}} & \multicolumn{3}{c}{} & \multicolumn{1}{c}{} & \multicolumn{1}{c}{} \\
    \cline{2-8}

    \multicolumn{1}{c|}{} & \multicolumn{2}{c|}{\textbf{Copilot}} & \multicolumn{1}{c|}{\textbf{Extension}} & \multicolumn{4}{c}{\textbf{Web Interface}} & \multicolumn{1}{|c}{} \\
    \hline

        \textbf{Prompt \#}  & \multicolumn{1}{c|}{\textbf{Claude}} & \multicolumn{1}{c|}{\textbf{ChatGPT}} & \multicolumn{1}{c|}{\textbf{Amazon Q}} & \multicolumn{1}{c|}{\textbf{Gemini}}& \multicolumn{1}{c|}{\textbf{ChatGPT}} & \multicolumn{1}{c|}{\textbf{DeepSeek}}  & 
        \multicolumn{1}{c|}{\textbf{Claude}} &
        \multicolumn{1}{c|}{\textbf{Average}} \\ \hline
        \rowcolor{prompt1}\textbf{\#1} & 63.20\% & 53.68\% & 56.99\% & 47.62\% & 67.53\% & 70.56\% & 76.77\%
        & \cellcolor{prompt1!40} 62.34\% \\ \hline
        
        \rowcolor{prompt2}\textbf{\#2} & 71.52\% & 79.39\% & 76.36\% & 41.82\% & 61.21\% & 82.42\% & \textbf{84.85}\% 
        & \cellcolor{prompt2!40} \textbf{71.08}\% \\ \hline
        
        \rowcolor{prompt3}\textbf{\#3 / Correct Programs} & 43.03\% & 51.52\% & 46.06\% & 45.45\% & 46.06\% & 76.97\% & 71.72\% 
        & \cellcolor{prompt3!40} \textbf{54.40}\% \\ \hline
        
        \rowcolor{prompt4}\textbf{\#4 / Correct Programs} & 63.03\% & 72.73\% & 65.45\% & \textbf{40.00}\% & 70.91\% & 77.58\% & 83.03\% 
        & \cellcolor{prompt4!40} 67.53\% \\ \hline
        \rowcolor{gray!10} \textbf{Average} & 60.19\% & 64.33\% & 61.22\% & \textbf{43.72}\% & 61.43\% & 76.88\% & \textbf{79.09}\%  \\
        \cline{1-8}

        \end{tabular}}
        \caption{Performance of all chatbots on the class of semantically correct programs across all prompts. Prompts 1 and 2 contain only correct implementations. To ensure a fair comparison, for Prompt 3 and Prompt 4 we report only the performance on the class of correct programs, \ie those that are truly semantically equivalent. The best and worst performances are highlighted in bold.}
        \label{tab:accuracy-correct-only}
        \vspace{-.5cm}
\end{table}

We also investigate the behavior of the chatbots when we insert in the prompt both correct and incorrect perturbations of the reference function (yellow and green rows). In this case, we observe a significant drop in performance compared to the single-class prompt (red and purple rows). Specifically, there is about an 8\% drop in average performance between Prompt 1 and Prompt 3, with DeepSeek being the only exception. The same phenomenon happens when we add context, as there is a degradation of about 3.5\% on average in performance between Prompt~2 and Prompt~4, with only ChatGPT seeming unaffected.

Finally, we report the overall results of the multi-class prompts for both classes and the partial results. Interestingly, \Cref{tab:accuracy-prompt3_4} exhibits good results in detecting incorrect implementations. In particular, if we compare the two tables, there is no real improvement between the contextless Prompt 3 and the contextual Prompt 4 concerning the class of incorrect programs. However, this increase comes at the expense of the class of semantically equivalent programs. The experiments thus show that there is a classification bias, where the chatbots tend to misclassify correct perturbations when they are also provided with incorrect implementations.

\Cref{tab:results_perturb} in the Appendix provides a detailed breakdown of the results for each type of code perturbation.  
For each of the 11 algorithms, the accuracy of all chatbots is reported across the three perturbations: copy propagation (CP), constant folding (CF), and their combination (CP + CF).  
Claude Sonnet and DeepSeek once again stand out, achieving the highest performances even on individual perturbation types.  
Conversely, Gemini performs worse than the other models overall.  
On average, constant folding is the least confusing perturbation for chatbots, yielding an accuracy of 76.40\%.
Copy propagation appears more challenging, with an accuracy of 69.20\%, while the combination of both perturbations results in the lowest accuracy, at 49.39\%.  

\begin{table}[t]
    \centering
    \setlength{\tabcolsep}{1pt} 
    \renewcommand{\arraystretch}{1.8} 
    
    \resizebox{\textwidth}{!}{
        \begin{tabular}{|p{4.5cm}|>{\raggedleft\arraybackslash}p{1.8cm}|>{\raggedleft\arraybackslash}p{1.8cm}|>{\raggedleft\arraybackslash}p{1.8cm}|>{\raggedleft\arraybackslash}p{1.8cm}|>{\raggedleft\arraybackslash}p{1.8cm}|>{\raggedleft\arraybackslash}p{1.8cm}|>{\raggedleft\arraybackslash}p{1.8cm}|>
    {\raggedleft\arraybackslash}p{1.8cm}|}
    \cline{2-4}

    \multicolumn{1}{c|}{} & \multicolumn{3}{c|}{\textbf{VSCode Plugin}} & \multicolumn{3}{c}{} & \multicolumn{1}{c}{} & \multicolumn{1}{c}{} \\
    \cline{2-8}

    \multicolumn{1}{c|}{} & \multicolumn{2}{c|}{\textbf{Copilot}} & \multicolumn{1}{c|}{\textbf{Extension}} & \multicolumn{4}{c}{\textbf{Web Interface}} & \multicolumn{1}{|c}{} \\
    \hline

        \textbf{Prompt \#} & \multicolumn{1}{c|}{\textbf{Claude}} & \multicolumn{1}{c|}{\textbf{ChatGPT}} & \multicolumn{1}{c|}{\textbf{Amazon Q}} & \multicolumn{1}{c|}{\textbf{Gemini}}& \multicolumn{1}{c|}{\textbf{ChatGPT}} & \multicolumn{1}{c|}{\textbf{DeepSeek}}  & 
        \multicolumn{1}{c|}{\textbf{Claude}} &
        \multicolumn{1}{c|}{\textbf{Average}} \\ \hline

        \rowcolor{prompt3}\textbf{\#3 / Correct Programs}& \textbf{43.03}\% & 51.52\% & 46.06\% & 45.45\% & 46.06\% & 76.97\% & 71.72\% 
        & \cellcolor{prompt3!50}\textbf{54.40}\% \\ \hline
        
        \rowcolor{prompt3}\textbf{\#3 / Incorrect Programs} & 94.55\% & 89.55\% & 95.00\% & \textbf{95.91}\% & 90.00\% & 95.45\% & 93.94\%
        & \cellcolor{prompt3!50}\textbf{93.48}\% \\ \hline
        
        \rowcolor{prompt3!50}\textbf{\#3 / Overall} & 72.47\% & 73.25\% & 74.03\% & 74.29\% & \textbf{71.17}\% & \textbf{87.53}\% & 76.19\% 
        & \cellcolor{prompt3!50}75.56\% \\ \hline

        \rowcolor{prompt4}\textbf{\#4 / Correct Programs} & 63.03\% & 72.73\% & 65.45\% & \textbf{40.00}\% & 70.91\% & 77.58\% & 83.03\% 
        & \cellcolor{prompt4!50} \textbf{67.53}\% \\ \hline
        
        \rowcolor{prompt4}\textbf{\#4 / Incorrect Programs} & \textbf{98.64}\% & 85.45\% & 96.82\% & 93.64\% & 90.45\% & 89.55\% & 96.36\% 
        & \cellcolor{prompt4!50} \textbf{92.99}\% \\ \hline
        
        \rowcolor{prompt4!50}\textbf{\#4 / Overall} & 83.38\% & 80.00\% & 83.38\% & \textbf{70.65}\% & 82.08\% & 84.42\% & \textbf{90.65}\% 
        & \cellcolor{prompt4!50}82.08\% \\ \hline
        \cline{1-8}
    \end{tabular}}
        \caption{Complete results on the classification of both semantically correct and incorrect programs. A comparison is made between prompts with and without contexts.}
        \label{tab:accuracy-prompt3_4}
        \vspace{-.7cm}
\end{table}

\section{Discussion}
\label{sec:discussion}

Our investigation into the semantic equivalence of Python programs using Large Language Models (LLMs) has revealed a number of intriguing behavioral patterns and inconsistencies. Our observations highlighted the presence of contradictions, hallucinations, and limited understanding of certain coding patterns.
Notably, the models occasionally \textit{contradict themselves} during the evaluation process, stating something at the beginning of the output and landing on the opposite conclusion by the end.
For example, an initial affirmative response regarding the equivalence of two functions is sometimes later followed by a retraction or a confusing double negative — a behavior that suggests a superficial rather than robust internal reasoning process.
Furthermore, we observed that in many cases, when models initially judged two versions of a program to be non-equivalent (e.g., the “cp” and “cp+cf” perturbed versions), they later stated that the two were equivalent.
This flip-flopping not only highlights the internal instability of the models’ reasoning but also raises concerns about their ability to maintain logical consistency across the whole output. 


Another intriguing observation concerns response \textit{verbosity}. When we instructed models to be concise, their answers were often more erroneous. Though not included in our statistics, this suggests that extended outputs, reflecting the model’s chain of thought, play a crucial role in accurate reasoning. Limiting verbosity may truncate this process, increasing errors—a finding that merits further investigation into the link between output length and reasoning quality in program understanding.

Overall, copy propagation appears to be particularly challenging for most models, with the combination of copy propagation and constant folding being the most problematic case.
A closer look at the data suggests this confusion becomes more pronounced when non-numerical data types, such as lists and arrays, are involved. This is likely due to the models' limited understanding of the subtle semantics of Python's \textit{assignment operator}, which behaves differently depending on the data type: numeric types are copied, whereas arrays and lists are merely aliased by reference.
In other words, a statement like \texttt{B = A}, where \texttt{A} is an array, results in a new binding rather than an actual copy of \texttt{A}. This distinction appears to be a common source of confusion among the tested models. 
Notably, only Claude Sonnet and Deepseek demonstrated some ability to differentiate between references and true copies in cases of variable rebinding.

The Sieve of Eratosthenes serves as a particularly revealing example.
This algorithm combines arithmetic operations with list manipulation, yielding the lowest accuracy scores across all models.
We argue that this poor performance stems from the modulus operation in the if-guard, which appears to confuse most LLMs in determining when the list removal operation takes place in the then-branch.
The issue is further exacerbated in perturbed versions of the code, where the presence of copies amplifies the confusion, as models struggle to distinguish between actual list copies and mere references.

Despite the obfuscation of function names, all models rapidly recognized even lesser-known algorithms.
Interestingly, the antialiasing algorithm was frequently identified as an “upscaling” technique, even though it also performs downscaling to induce a blur effect.
In other words, none of the tested models fully grasped the antialiasing algorithm’s dual role — upscaling followed by downscaling — which constitutes a hallmark of its functionality.

Surprisingly, when tested on the 3D point rotation and FFT algorithms, all models achieved a 100\% success rate, suggesting that even less canonical snippets are well represented in their extensive knowledge base.
Conversely, the unification algorithm was consistently the least understood. 
This is likely due to the nature of the data structures involved: the algorithm computes a substitution — specifically, the Most General Unifier (MGU) — by performing pattern matching over pairs of types. 
Substitutions are essentially dictionaries, while types are syntactic entities defined through case-based structures in an object-oriented manner.
Such a use of classes appears to confuse chatbots, which seem better suited to understanding numerical algorithms, or those involving conventional data structures such as arrays and lists.


\subsection{Improving LLMs in Code Understanding}
\label{sec:improving}


LLMs' ability to understand code depends on the complexity of the chatbot's model ecosystem, where models may be dynamically selected based on the query type.  
For instance, ChatGPT appears to use different models/extensions depending on whether the task involves text, images, or code execution.  

To assess execution capabilities, we asked chatbots if they had direct access to a Python interpreter. All but two - ChatGPT and Gemini - denied this ability. ChatGPT consistently confirmed it could execute Python code, though OpenAI provides no official confirmation. Gemini, on the other hand, gave inconsistent answers, sometimes affirming and other times denying interpreter access, suggesting variable behavior or conditional access. 

The complexity of these ecosystems could readily accommodate the integration of code verification and analysis tools operating behind the scenes.
If chatbots incorporated a pre-processing phase with a pipeline of code transformation tools designed to enhance reasoning quality, then exploiting a code optimizer applying the rules of Section~\ref{sec:optimizations} could prove highly effective.

Fine-tuning \cite{latif2024fine,han2024parameter} offers another approach, where network weights are adjusted to incorporate new knowledge with existing information. In this context, we could invert the transformation rules outlined in \Cref{sec:optimizations}, using a dataset of both original and perturbed code snippets for training. However, fine-tuning requires careful selection of a large sample of code, such as for constant folding, and is limited by the proprietary nature of the analyzed chatbots, meaning only their respective owners can perform such training, restricting academic involvement.

\section{Conclusions}
\label{sec:conclusions}

We analyzed LLMs' code understanding by integrating insights from compiler design and advances in generative AI.  
To generate semantically equivalent code requiring nuanced comprehension, we employed standard compiler optimization techniques: copy propagation and constant folding.  
Since these optimizations are typically defined on control flow graphs of intermediate representations, we adapted them to work directly at the source code level.  

Our experiments highlight both the strengths and limitations of current LLMs in code understanding.  
While contextual prompting can improve performance in some cases, it also reveals inconsistencies in logical reasoning.  
Results indicate that zero-shot contextless queries yield a 59\% accuracy, while adding contextual information improves to a 71\%.  
Despite a 12\% increase, a 29\% error rate is still too high to broadly trust these models for code reasoning.
Our suggestion is that a code pre-processing tool integrated into the LLM pipeline would improve program understanding by a fair amount.

Future work should disentangle these effects, refining both LLM reasoning and evaluation methods for complex semantic tasks.  
Further directions include expanding the dataset, testing additional models, and exploring more advanced optimizations.

\bibliographystyle{splncs04}
\bibliography{references}
\newpage

\appendix

\section{Inference Rules for Copy Propagation and Constant Folding}

This appendix collects the technical material for annotating {\Python} programs with relevant
informations that support copy propagation and constant folding. We focus on a fundamental subset of {\Python}  that remains expressive enough to encompass all the tests in \Cref{sec:experimental-setup}.
The extension of the techniques to the whole language is out of the scope
of this paper (and will be addressed in a future work).
To make the Appendix self-contained, when necessary, we will recall the notations we use.

\subsection{Copy Propagation Analysis}
Let $\Var(\P)$ be the set of variables of a {\Python} program {\P}; we also apply $\Var(\cdot)$
to statements $S$ with the same meaning. 
The annotations of copy propagation are sets $\Pset$ of pairs $(x,y)$, 
with $x, y \in \Var(\P)$, that are closed by symmetry and transitivity; the set $\{(x,y), (y,x)\}$ is 
abbreviated into $x \newsim y$ in the
following. Let
\begin{itemize}
\item
$\Pset \setminus x \eqdef \{ (y,z) \in \Pset \; | \; y \neq x \; \text{ and } \; z \neq x \}$ ;
\item
$\Pset \cap \Pset' \eqdef \{ (x,y) \; | \; (x,y) \in \Pset \; \text{ and } \; (x,y) \in \Pset' \}$;
\item
$(A)^{\tt st}$ is the closure of a set $A$ by symmetry and transitivity.
\end{itemize}

We use \emph{judgments} of the form  %
$\jexp{\Pset}{S}{\Pset'}$ for binding statements to annotations $\Pset$ meaning that, if $S$ is typed with an 
initial set of copies $\Pset$,
it will produce a set of copies $\Pset'$. The rules for deriving copies of a {\Python} program $\P$ 
are written in Table~\ref{tab:typeinferenceCP}. 
Few remarks about the most relevant rules follow. Rule~\rulename{Asgn-CP} defines the annotation of an
assignment to a variable $x$. In this case ($E$ is not a variable), any copy of $x$ is broken and must be 
removed from the set
$\Pset$, hence the conclusion $\Pset\setminus x$. Rule \rulename{Id-CP} destroy previous copies of
$x$ and creates a copy with $y$ and with all the variables that are copies of $y$ in $\Pset$.
Rule~\rulename{While-CP} is the critical one. To determine the set $\Pset''$ such that 
$\jexp{\Pset}{\while{E}{S}}{\Pset''}$ we need an invariant: a set $\Pset'$ such that 
$\jexp{\Pset'}{S}{\Pset'}$. However, a generic set $\Pset'$ in not correct because we cannot have more 
copies then those when the control reaches the first time the iteration. Hence the invariant becomes
$\jexp{\Pset \cap \Pset'}{S}{\Pset'}$ and the conclusion of \rulename{While-CP} takes into account this 
invariant and the fact that the iteration may be not executed at all. Rule~\rulename{For-CP} is similar 
to \rulename{While-CP}. Rules~\rulename{Def-CP} 
and~\rulename{Call-CP} deal with function definitions and invocations; they assume that no access to global
variables is possible. With this constraint, the judgments are pretty easy.

The main difficulty of the system in Table~\ref{tab:typeinferenceCP} is determining the invariants 
in correspondence of iterations. There is a standard way to solve this issue: starting with empty 
annotations, running the type systems several times until the annotations do not change anymore. This 
algorithm always terminates (given a program, the corresponding sets $\Pset$ are a finite lattice
and the iterations always return greater annotations). 
\begin{table}[t]
{\small \[
\begin{array}{c}
\mathrule{Asgn-CP}{
E \; \text{ \emph{is not a variable}}
}{
	\jexp{\Pset}{x = E}{\Pset \setminus x}
	}
\quad
\mathrule{Id-CP}{
x \neq y 
}{
	\jexp{\Pset}{x = y}{(\Pset \setminus x \cup \{ (x,y), (y,x) \})^{\tt st}}
	}
\quad
\mathax{Ids-CP}{
	\jexp{\Pset}{x = x}{\Pset}
	}
\\
\\
\mathrule{If-CP}{
	\jexp{\Pset}{S}{\Pset'} \qquad \jexp{\Pset}{S'}{\Pset''}
	}{
	\jexp{\Pset}{\ifte{E}{S}{S'}}{\Pset' \cap \Pset''}
	}
\qquad
\mathrule{While-CP}{
	\jexp{\Pset \cap \Pset'}{S}{\Pset'} 
	}{
	\jexp{\Pset}{\while{E}{S}}{\Pset \cap \Pset'}
	}
\\
\\
\mathrule{For-CP}{
	\jexp{\Pset \cap \Pset'\setminus i}{S}{\Pset'} 
	}{
	\jexp{\Pset}{\for{i}{E}{S}}{\Pset \cap \Pset'\setminus i}
	}
\qquad
\mathrule{Seq-CP}{
	\jexp{\Pset}{S}{\Pset'} \qquad \jexp{\Pset'}{S'}{\Pset''} 
	}{
		\jexp{\Pset}{S \; \; S'}{\Pset''}
	}
\\
\\
\qquad
\mathrule{Def-CP}{
	\jexp{\emptyset}{S}{\Pset'}
	}{
		\jexp{\Pset}{\deffun{f}{x_1, \cdots, x_n}{S}}{\Pset}
	}
\qquad
\mathax{Call-CP}{
		\jexp{\Pset}{f(e_1, \cdots, e_n)}{\Pset}
	}
\end{array}
\]}
\caption{\label{tab:typeinferenceCP} Rewriting rules for Copy Propagation.}
\end{table}
As an example, we apply the algorithm to a simple code printing the factorial of a value
taken in input. In the following table, we abbreviate $\{ (x,y), (y,x)\}$ with $x \sim y$ and,
instead of writing 
$\jexp{\Pset}{S}{\Pset'}$, we write $S \; \; \Pset, \Pset'$).
\begin{center}{\small
\begin{tabular}{l@{\qquad}l@{\quad}l@{\quad}l}
code & approximant 1 & approximant 2 & approximant 3
\\
\\
{\tt n = int(input())} & $\varnothing, \varnothing$ & $\varnothing, \varnothing$ & $\varnothing, \varnothing$
\\
{\tt m = 1} & $\varnothing, \varnothing$ & $\varnothing, \varnothing$ & $\varnothing, \varnothing$
\\
{\tt tmp = n ;} & $\varnothing, \varnothing$ & $\varnothing, {\tt n} \sim {\tt tmp}$ & $\varnothing, {\tt n} \sim {\tt tmp}$
\\
{\tt while (n>1):}& $\varnothing, \varnothing$ & $\varnothing, \varnothing$ & ${\tt n} \sim {\tt tmp}, {\tt n} \sim {\tt tmp}$
\\
\qquad {\tt m = m * n}& $\varnothing, \varnothing$ & $\varnothing, \varnothing$ & ${\tt n} \sim {\tt tmp}, {\tt n} \sim {\tt tmp}$
\\
\qquad {\tt tmp = n - 1}& $\varnothing, \varnothing$ & $\varnothing, \varnothing$ & ${\tt n} \sim {\tt tmp}, \varnothing$
\\
\qquad {\tt n = tmp}& $\varnothing, \varnothing$ & $\varnothing, {\tt n} \sim {\tt tmp}$ & $\varnothing, {\tt n} \sim {\tt tmp}$
\\
{\tt print(m)}& $\varnothing, \varnothing$ & $\varnothing, \varnothing$ & ${\tt n} \sim {\tt tmp}, {\tt n} \sim {\tt tmp}$
\end{tabular}
}\end{center}
In this case, the algorithm terminates after three applications of the type system in 
Table~\ref{tab:typeinferenceCP} -- the three approximants in the above picture. 

\subsection{Constant Folding}
We use \emph{abstract memories} $\Cset$ that are sets of pairs $(x, \mu)$ where $x \in \Var(\P)$ (for some 
{\Python} program $\P$) and $\mu$ is either a constant value or $\top$, where $\top$ represents 
a non-constant value, or \mbox{\tt ?} that represents an error. We have the following operations:
\begin{itemize}
\item
$\Cset \setminus x \eqdef \{ (y,\mu) \in \Cset \; | \; y \neq x \}$ ;

\medskip

\item
$\mu \sqcup \mu' \eqdef \left\{
\begin{array}{l@{\quad}l}
\mu & \text{if } \; \mu = \mu'
\\
\top & \text{if } \; \mu \neq \mu' \; \text{ and } \; \mu, \mu' \neq \mbox{\tt ?} 
\\
\mbox{\tt ?} & \text{if } \; \mu = \mbox{\tt ?} \; \text{ or } \; \mu' = \mbox{\tt ?} 
\end{array}\right.
$
\item
$\Cset \sqcup \Cset' \eqdef 
\left\{ 
\begin{array}{l@{\qquad}l}
\{ (x, \mu \sqcup \mu') \} \cup (\Cset \setminus x \sqcup \Cset' \setminus x) & 
\text{if } \; (x,\mu) \in \Cset  \; \text{ and } \; (x,\mu') \in \Cset' 
\\
\{ (x, \mu) \} \cup (\Cset \setminus x \sqcup \Cset') & 
\text{if } \; (x,\mu) \in \Cset  \; \text{ and } \; \Cset'\setminus x = \Cset' 
\\
\{ (x, \mu') \} \cup (\Cset  \sqcup \Cset'\setminus x) & 
\text{if } \; (x,\mu') \in \Cset'  \; \text{ and } \; \Cset\setminus x = \Cset 
\end{array}
\right.
$

\medskip

\item
$\sem{E}_{\Cset}$ that evaluates $E$ in the abstract memory $\Cset$; if some variable in $E$ has value 
{\tt ?} in $\Cset$ or has no value then the result is {\tt ?} (error); otherwise if some variable is $\top$ 
(and no variable is {\tt ?} or has no value in $\Cset$) 
then the overall result is $\top$;
otherwise the result is the value of $E$ when all the variables have been replaced by the corresponding values in 
$\Cset$ (errors may occur at this stage as well, \emph{e.g.}~the division by 0 returns {\tt ?}).
\end{itemize}

The \emph{judgments} for the constant folding are
$\jexp{\Cset}{S}{\Cset'}$, meaning that, if $S$ is typed with an 
initial abstract memory $\Cset$, then its evaluation terminates with a memory whose variables have the values stored in $\Cset'$. 
The rules for deriving copies of a {\Python} program $\P$ 
are written in Table~\ref{tab:typeinferenceCF}. 
\begin{table}[t]
{\small \[
\begin{array}{c}
\mathrule{Asgn-CF}{
	\sem{E}_{\Cset} = \mu
	}{
	\jexp{\Cset}{x = E}{\Cset \setminus x \cup \{ (x, \mu) \}}
	}
\qquad
\mathrule{If-CF}{
	\sem{E}_{\Cset} = \top \quad \jexp{\Cset}{S}{\Cset'} \quad \jexp{\Cset}{S'}{\Cset''}
	}{
	\jexp{\Cset}{\ifte{E}{S}{S'}}{\Cset' \sqcup \Cset''}
	}
\\
\\
\mathrule{If-CF-t}{
	\sem{E}_{\Cset} = {\it true} \quad \jexp{\Cset}{S}{\Cset'} 
	}{
	\jexp{\Cset}{\ifte{E}{S}{S'}}{\Cset'}
	}
\qquad
\mathrule{If-CF-f}{
	\sem{E}_{\Cset} = {\it false} \quad \jexp{\Cset}{S'}{\Cset''}
	}{
	\jexp{\Cset}{\ifte{E}{S}{S'}}{\Cset''}
	}
\\
\\
\mathrule{While-CF}{
	\sem{E}_{\Cset} = \mu \quad \mu \neq {\it false} \quad \jexp{\Cset \sqcup \Cset'}{S}{\Cset'} 
	}{
	\jexp{\Cset}{\while{E}{S}}{\Cset \sqcup \Cset'}
	}
\qquad
\mathrule{While-CF-f}{
	\sem{E}_{\Cset} = {\it false} \quad \jexp{\Cset \sqcup \Cset'}{S}{\Cset'} 
	}{
	\jexp{\Cset}{\while{E}{S}}{\Cset}
	}
\\
\\
\mathrule{For-CF}{
	\begin{array}{c}
	\sem{E}_{\Cset} = \mu \quad (\mu > 0 \text{ or } \mu = \top)
	\\
	\jexp{\Cset \sqcup \Cset' \sqcup \{(i,\top)\}}{S}{\Cset'} 
	\end{array}
	}{
	\jexp{\Cset}{\for{i}{E}{S}}{\Cset \sqcup \Cset' \sqcup \{(i,\top)\}}
	}
\quad
\mathrule{For-CF-f}{
	\sem{E}_{\Cset} = {\tt k} \quad {\tt k} \leq 0 
	}{
	\jexp{\Cset}{\for{i}{E}{S}}{\Cset}
	}
\\
\\
\mathrule{Seq-CF}{
	\jexp{\Cset}{S}{\Cset'} \quad \jexp{\Cset'}{S'}{\Cset''} 
	}{
		\jexp{\Cset}{S \; \; S'}{\Cset''}
	}
\quad
\mathrule{Def-CF}{
	\jexp{\top}{S}{\Cset'}
	}{
		\jexp{\Cset}{\deffun{f}{x_1, \cdots, x_n}{S}}{\Cset}
	}
\quad
\mathax{Call-CF}{
		\jexp{\Cset}{f(e_1, \cdots, e_n)}{\Cset}
	}
\end{array}
\]}
\caption{Rewriting rules for Constant Folding.}
\label{tab:typeinferenceCF} 
\end{table}
We comment rules~\rulename{Asgn-CF} and~\rulename{If-CF}, the comments for the other ones are similar 
to those of the copy propagation. Rule~\rulename{Asgn-CF} updates the abstract memory $\Cset$  
by binding $x$ with the value $\sem{E}_\Cset$. Rule~\rulename{If-CF} types conditionals when the
value of the guard is not determined. In this case the abstract memory after the conditional must include
the results $\Cset'$ and $\Cset''$ of the two branches (at static time we are not aware of the branches 
that will be executed). Hence the abstract memory $\Cset' \sqcup \Cset''$ is the conclusion.

Like the copy propagation analysis, the constant folding requires an iterative algorithm to
determine the annotations. The algorithm starts with empty abstract memories. 
As an example, we apply the algorithm to a simple code printing the factorial of a value
taken in input:
\begin{center}{\small
\begin{tabular}{l@{\qquad}l@{\quad}l@{\quad}l}
code & approximant 1 & approximant 2 & approximant 3
\\[.3cm]
{\tt n = int(input())} & $\emptyset, \; \emptyset$ & $\emptyset, \; \{({\tt n},\top)\}$ & $\emptyset, \; \{({\tt n},\top)\}$
\\
{\tt tmp = 1} & $\emptyset, \;\emptyset$ & $\{({\tt n},\top)\}, \;\{({\tt n},\top), ({\tt tmp},1) \}$ & $\{({\tt n},\top)\}, \;\{({\tt n},\top), ({\tt tmp},1) \}$
\\
{\tt m = 2*tmp - 1} & $\emptyset, \; \emptyset$ & $\{({\tt n},\top), ({\tt tmp},1) \}, \; \Cset_1$ & 
$\{({\tt n},\top), ({\tt tmp},1) \}, \; \Cset_1$
\\
{\tt while (n > tmp):}& $\emptyset, \; \emptyset$ & $\Cset_1, \; \Cset_1$ & $\Cset_4, \; \Cset_4$
\\
\qquad {\tt tmp = tmp + 1}& $\emptyset, \; \emptyset$ & $\Cset_1, \;\Cset_2$ & $\Cset_4, \;\Cset_3$
\\
\qquad {\tt m = m * n}& $\emptyset, \; \emptyset$ & $\Cset_2, \; \Cset_3$ & $\Cset_3, \;\Cset_3$
\\
\qquad {\tt n = n - tmp + 1}& $\emptyset, \; \emptyset$ & $\Cset_3, \; \Cset_3$ & $\Cset_3, \;\Cset_3$
\\
\qquad {\tt tmp = tmp - 1} & $\emptyset, \; \emptyset$ & $\Cset_3, \; \Cset_4$ & $\Cset_3, \;\Cset_4$
\\
{\tt print(m)}& $\emptyset, \; \emptyset$ & $\Cset_4,\; \Cset_4$ & $\Cset_4, \;\Cset_4$ 
\end{tabular}
}\end{center}
where
\[
\begin{array}{c}
\Cset_1 = \{({\tt n},\top), ({\tt tmp},1), ({\tt m},1)  \} \qquad 
\Cset_2 = \{({\tt n},\top), ({\tt tmp},2), ({\tt m},1)  \}
\\
\Cset_3 = \{({\tt n},\top), ({\tt tmp},2), ({\tt m},\top)  \} \qquad
\Cset_4 = \{({\tt n},\top), ({\tt tmp},1), ({\tt m},\top)  \} \; .
\end{array}
\]
Also in this case, the algorithm terminates after three applications of the type system in 
Table~\ref{tab:typeinferenceCF}. 

\section{Performance evaluation aggregated by the type of perturbation}
The following table presents aggregated results for each algorithm in the dataset along with the corresponding semantically preserving perturbations.

\begin{table}[t]
    \centering
    \setlength{\tabcolsep}{2.5pt} 
    \renewcommand{\arraystretch}{1.8} 
    \resizebox{\textwidth}{!}{
        \begin{tabular}{|>{\centering\arraybackslash}p{3.2cm}|>{\centering\arraybackslash}p{1.6cm}|>{\raggedleft\arraybackslash}p{1.8cm}|>{\raggedleft\arraybackslash}p{1.8cm}|>{\raggedleft\arraybackslash}p{1.8cm}|>{\raggedleft\arraybackslash}p{1.8cm}|>{\raggedleft\arraybackslash}p{1.8cm}|>{\raggedleft\arraybackslash}p{1.8cm}|>
    {\raggedleft\arraybackslash}p{1.8cm}|>{\raggedleft\arraybackslash}p{1.8cm}|}
    \cline{3-5}
    \multicolumn{2}{c|}{} & \multicolumn{3}{c|}{\textbf{VSCode Plugin}} & \multicolumn{3}{c}{} & \multicolumn{1}{c}{} & \multicolumn{1}{c}{} \\
    \cline{3-9}
    \multicolumn{2}{c|}{} & \multicolumn{2}{c|}{\textbf{Copilot}} & \multicolumn{1}{c|}{\textbf{Extension}} & \multicolumn{4}{c}{\textbf{Web Interface}} & \multicolumn{1}{|c}{} \\
    \cline{2-10}
        \multicolumn{1}{c|}~ & \textbf{Perturb} & \multicolumn{1}{c|}{\textbf{Claude}} & \multicolumn{1}{c|}{\textbf{ChatGPT}} & \multicolumn{1}{c|}{\textbf{Amazon Q}} & \multicolumn{1}{c|}{\textbf{Gemini}}& \multicolumn{1}{c|}{\textbf{ChatGPT}} & \multicolumn{1}{c|}{\textbf{DeepSeek}}  & 
        \multicolumn{1}{c|}{\textbf{Claude}} &
        \multicolumn{1}{c|}{\textbf{Average}} \\ \hline

        \multirow{3}{*}{\textbf{Duplicate Removal}} & \cellcolor{lg} cp  & \cellcolor{lg} 30.00\% & \cellcolor{lg} 85.71\% & \cellcolor{lg} 49.29\% & \cellcolor{lg} 5.00\% & \cellcolor{lg} 86.43\% & \cellcolor{lg} 80.00\% & \cellcolor{lg} 100.00\% & \cellcolor{lg} 62.35\% \\
        ~ & \cellcolor{mg} cf & \cellcolor{mg} 82.50\% & \cellcolor{mg} 89.29\% & \cellcolor{mg} 95.00\% & \cellcolor{mg} 40.00\% & \cellcolor{mg} 95.00\% & \cellcolor{mg} 100.00\% & \cellcolor{mg} 100.00\% & \cellcolor{mg} 85.97\% \\ 
        ~ & \cellcolor{dg} cp+cf & \cellcolor{dg} 40.00\% & \cellcolor{dg} 60.71\% & \cellcolor{dg} 44.29\% & \cellcolor{dg} 0.00\% & \cellcolor{dg} 91.43\% & \cellcolor{dg} 100.00\% & \cellcolor{dg} 75.00\% & \cellcolor{dg} 58.78\% \\ \hline

        \multirow{3}{*}{\textbf{Sieve}} & \cellcolor{lg} cp & \cellcolor{lg} 17.50\% & \cellcolor{lg} 35.00\% & \cellcolor{lg} 5.00\% & \cellcolor{lg} 14.29\% & \cellcolor{lg} 8.57\% & \cellcolor{lg} 18.57\% & \cellcolor{lg} 8.33\% & \cellcolor{lg} \textbf{15.32\%} \\ 
        ~ & \cellcolor{mg} cf & \cellcolor{mg} 35.00\% & \cellcolor{mg} 70.00\% & \cellcolor{mg} 30.00\% & \cellcolor{mg} 3.57\% & \cellcolor{mg} 30.00\% & \cellcolor{mg} 62.14\% & \cellcolor{mg} 100.00\% & \cellcolor{mg} 47.24\% \\ 
        ~ & \cellcolor{dg} cp+cf & \cellcolor{dg} 10.00\% & \cellcolor{dg} 20.00\% & \cellcolor{dg} 0.00\% & \cellcolor{dg} 0.00\% & \cellcolor{dg} 0.00\% & \cellcolor{dg} 8.57\% & \cellcolor{dg} 8.33\% & \cellcolor{dg} \textbf{6.70\%} \\ \hline

        \multirow{3}{*}{\textbf{Count Occurrences}}  & \cellcolor{lg} cp & \cellcolor{lg} 60.00\% & \cellcolor{lg} 100.00\% & \cellcolor{lg} 52.14\% & \cellcolor{lg} 19.29\% & \cellcolor{lg} 62.86\% & \cellcolor{lg} 91.43\% & \cellcolor{lg} 82.50\% & \cellcolor{lg} 66.89\% \\ 
        ~ & \cellcolor{mg} cf & \cellcolor{mg} 92.50\% & \cellcolor{mg} 100.00\% & \cellcolor{mg} 100.00\% & \cellcolor{mg} 67.86\% & \cellcolor{mg} 70.00\% & \cellcolor{mg} 95.00\% & \cellcolor{mg} 100.00\% & \cellcolor{mg} 89.34\% \\ 
        ~ & \cellcolor{dg} cp+cf & \cellcolor{dg} 57.50\% & \cellcolor{dg} 75.00\% & \cellcolor{dg} 52.14\% & \cellcolor{dg} 15.71\% & \cellcolor{dg} 66.43\% & \cellcolor{dg} 95.00\% & \cellcolor{dg} 82.50\% & \cellcolor{dg} 63.47\% \\ \hline
        
        \multirow{3}{*}{\textbf{Fibonacci}} & \cellcolor{lg} cp & \cellcolor{lg} 75.00\% & \cellcolor{lg} 58.57\% & \cellcolor{lg} 85.71\% & \cellcolor{lg} 45.71\% & \cellcolor{lg} 47.86\% & \cellcolor{lg} 86.43\% & \cellcolor{lg} 100.00\% & \cellcolor{lg} 71.33\% \\ 
        ~ & \cellcolor{mg} cf & \cellcolor{mg} 87.50\% & \cellcolor{mg} 96.43\% & \cellcolor{mg} 96.43\% & \cellcolor{mg} 47.14\% & \cellcolor{mg} 57.86\% & \cellcolor{mg} 96.43\% & \cellcolor{mg} 100.00\% & \cellcolor{mg} 83.11\% \\ 
        ~ & \cellcolor{dg} cp+cf & \cellcolor{dg} 30.00\% & \cellcolor{dg} 5.00\% & \cellcolor{dg} 37.14\% & \cellcolor{dg} 10.00\% & \cellcolor{dg} 29.29\% & \cellcolor{dg} 27.14\% & \cellcolor{dg} 70.00\% & \cellcolor{dg} 29.80\% \\ \hline
        
        \multirow{3}{*}{\textbf{Primality}} & \cellcolor{lg} cp & \cellcolor{lg} 20.00\% & \cellcolor{lg} 33.57\% & \cellcolor{lg} 25.00\% & \cellcolor{lg} 7.14\% & \cellcolor{lg} 62.86\% & \cellcolor{lg} 95.00\% & \cellcolor{lg} 49.17\% & \cellcolor{lg} 41.82\% \\ 
        ~ & \cellcolor{mg} cf & \cellcolor{mg} 87.50\% & \cellcolor{mg} 87.86\% & \cellcolor{mg} 96.43\% & \cellcolor{mg} 72.14\% & \cellcolor{mg} 90.00\% & \cellcolor{mg} 90.00\% & \cellcolor{mg} 100.00\% & \cellcolor{mg} 89.13\% \\ 
        ~ & \cellcolor{dg} cp+cf & \cellcolor{dg} 20.00\% & \cellcolor{dg} 25.00\% & \cellcolor{dg} 5.00\% & \cellcolor{dg} 3.57\% & \cellcolor{dg} 17.14\% & \cellcolor{dg} 15.00\% & \cellcolor{dg} 49.17\% & \cellcolor{dg} 19.27\% \\ \hline

        \multirow{3}{*}{\textbf{Find Duplicate}} & \cellcolor{lg} cp & \cellcolor{lg} 87.50\% & \cellcolor{lg} 95.00\% & \cellcolor{lg} 95.00\% & \cellcolor{lg} 100.00\% & \cellcolor{lg} 85.00\% & \cellcolor{lg} 100.00\% & \cellcolor{lg} 100.00\% & \cellcolor{lg} 94.64\% \\ 
        ~ & \cellcolor{mg} cf & \cellcolor{mg} 52.50\% & \cellcolor{mg} 77.86\% & \cellcolor{mg} 50.00\% & \cellcolor{mg} 15.71\% & \cellcolor{mg} 44.29\% & \cellcolor{mg} 95.00\% & \cellcolor{mg} 100.00\% & \cellcolor{mg} 62.19\% \\ 
        ~ & \cellcolor{dg} cp+cf & \cellcolor{dg} 37.50\% & \cellcolor{dg} 52.86\% & \cellcolor{dg} 50.00\% & \cellcolor{dg} 10.71\% & \cellcolor{dg} 39.29\% & \cellcolor{dg} 95.00\% & \cellcolor{dg} 100.00\% & \cellcolor{dg} 55.05\% \\ \hline
        
        \multirow{3}{*}{\textbf{Bubble sort}} & \cellcolor{lg} cp & \cellcolor{lg} 37.50\% & \cellcolor{lg} 44.29\% & \cellcolor{lg} 37.86\% & \cellcolor{lg} 50.00\% & \cellcolor{lg} 47.86\% & \cellcolor{lg} 58.57\% & \cellcolor{lg} 82.50\% & \cellcolor{lg} 51.22\% \\ 
        ~ & \cellcolor{mg} cf & \cellcolor{mg} 80.00\% & \cellcolor{mg} 92.86\% & \cellcolor{mg} 85.00\% & \cellcolor{mg} 81.43\% & \cellcolor{mg} 96.43\% & \cellcolor{mg} 100.00\% & \cellcolor{mg} 100.00\% & \cellcolor{mg} 90.82\% \\ 
        ~ & \cellcolor{dg} cp+cf & \cellcolor{dg} 37.50\% & \cellcolor{dg} 60.71\% & \cellcolor{dg} 22.86\% & \cellcolor{dg} 46.43\% & \cellcolor{dg} 40.71\% & \cellcolor{dg} 58.57\% & \cellcolor{dg} 82.50\% & \cellcolor{dg} 49.90\% \\ \hline
        
        \multirow{3}{*}{\textbf{Anti Aliasing}} & \cellcolor{lg} cp & \cellcolor{lg} 92.50\% & \cellcolor{lg} 70.71\% & \cellcolor{lg} 95.00\% & \cellcolor{lg} 55.71\% & \cellcolor{lg} 81.43\% & \cellcolor{lg} 89.29\% & \cellcolor{lg} 95.83\% & \cellcolor{lg} 82.93\% \\ 
        ~ & \cellcolor{mg} cf & \cellcolor{mg} 30.00\% & \cellcolor{mg} 75.71\% & \cellcolor{mg} 20.00\% & \cellcolor{mg} 69.29\% & \cellcolor{mg} 46.43\% & \cellcolor{mg} 100.00\% & \cellcolor{mg} 86.67\% & \cellcolor{mg} 61.16\% \\ 
        ~ & \cellcolor{dg} cp+cf & \cellcolor{dg} 15.00\% & \cellcolor{dg} 80.71\% & \cellcolor{dg} 10.00\% & \cellcolor{dg} 27.14\% & \cellcolor{dg} 46.43\% & \cellcolor{dg} 70.71\% & \cellcolor{dg} 62.50\% & \cellcolor{dg} 44.64\% \\ \hline
        
        \multirow{3}{*}{\textbf{FFT}} & \cellcolor{lg} cp & \cellcolor{lg} 92.50\% & \cellcolor{lg} 100.00\% & \cellcolor{lg} 100.00\% & \cellcolor{lg} 100.00\% & \cellcolor{lg} 95.00\% & \cellcolor{lg} 100.00\% & \cellcolor{lg} 100.00\% & \cellcolor{lg} 98.21\% \\ 
        ~ & \cellcolor{mg} cf & \cellcolor{mg} 92.50\% & \cellcolor{mg} 100.00\% & \cellcolor{mg} 100.00\% & \cellcolor{mg} 100.00\% & \cellcolor{mg} 95.00\% & \cellcolor{mg} 100.00\% & \cellcolor{mg} 100.00\% & \cellcolor{mg} 98.21\% \\ 
        ~ & \cellcolor{dg} cp+cf & \cellcolor{dg} 92.50\% & \cellcolor{dg} 100.00\% & \cellcolor{dg} 100.00\% & \cellcolor{dg} 95.00\% & \cellcolor{dg} 85.00\% & \cellcolor{dg} 100.00\% & \cellcolor{dg} 100.00\% & \cellcolor{dg} 96.07\% \\ \hline
        
        \multirow{3}{*}{\textbf{Rotate 3D}} & \cellcolor{lg} cp & \cellcolor{lg} 92.50\% & \cellcolor{lg} 100.00\% & \cellcolor{lg} 100.00\% & \cellcolor{lg} 100.00\% & \cellcolor{lg} 100.00\% & \cellcolor{lg} 100.00\% & \cellcolor{lg} 100.00\% & \cellcolor{lg} 98.93\% \\ 
        ~ & \cellcolor{mg} cf & \cellcolor{mg} 75.00\% & \cellcolor{mg} 100.00\% & \cellcolor{mg} 100.00\% & \cellcolor{mg} 85.00\% & \cellcolor{mg} 100.00\% & \cellcolor{mg} 100.00\% & \cellcolor{mg} 100.00\% & \cellcolor{mg} 94.29\% \\ 
        ~ & \cellcolor{dg} cp+cf & \cellcolor{dg} 62.50\% & \cellcolor{dg} 85.00\% & \cellcolor{dg} 100.00\% & \cellcolor{dg} 70.00\% & \cellcolor{dg} 100.00\% & \cellcolor{dg} 100.00\% & \cellcolor{dg} 86.67\% & \cellcolor{dg} 86.31\% \\ \hline
        
        \multirow{3}{*}{\textbf{Unification}} & \cellcolor{lg} cp & \cellcolor{lg} 87.50\% & \cellcolor{lg} 75.00\% & \cellcolor{lg} 96.43\% & \cellcolor{lg} 75.00\% & \cellcolor{lg} 68.57\% & \cellcolor{lg} 63.57\% & \cellcolor{lg} 76.67\% & \cellcolor{lg} 77.53\% \\ 
        ~ & \cellcolor{mg} cf & \cellcolor{mg} 47.50\% & \cellcolor{mg} 45.00\% & \cellcolor{mg} 46.43\% & \cellcolor{mg} 10.00\% & \cellcolor{mg} 20.00\% & \cellcolor{mg} 22.14\% & \cellcolor{mg} 81.67\% & \cellcolor{mg} 38.96\% \\ 
        ~ & \cellcolor{dg} cp+cf & \cellcolor{dg} 37.50\% & \cellcolor{dg} 45.00\% & \cellcolor{dg} 35.71\% & \cellcolor{dg} 0.00\% & \cellcolor{dg} 20.00\% & \cellcolor{dg} 23.57\% & \cellcolor{dg} 71.67\% & \cellcolor{dg} 33.35\% \\ \hline

    \end{tabular}}
    \caption{Aggregated results for each algorithtm in the dataset and the corresponding semantically-preserving perturbations.}\label{tab:results_perturb}
\end{table}

\end{document}